\documentclass[nopreprint,author-year,floats,floatfix,a4paper]{jasatex} 
\usepackage{microtype} 

\usepackage{verbatim}

\usepackage{amsmath,amssymb}
\usepackage[breaklinks]{hyperref} 

\usepackage{color}
\usepackage{soul}

\usepackage[normalem]{ulem}

\usepackage{srcltx}

\newcommand{\insfig}[1]{\includegraphics[width=\columnwidth]{#1}}

\newcommand{\Lerch}[3]{\ensuremath{\Phi}\!\left(#1,#2,#3\right)}

\newcommand{\dd}{\ensuremath{\text{d}}}

\newcommand{\eg}{\mbox{\emph{e.g.}}}

\newcommand{\taue}{\ensuremath{\tau_\epsilon}}
\newcommand{\taus}{\ensuremath{\tau_\sigma}}
\newcommand{\taun}{\ensuremath{\tau_\nu}}

\newcommand{\kappaz}{\ensuremath{\kappa_0}}
\newcommand{\kappan}{\ensuremath{\kappa_\nu}}

\newcommand{\OmegaI}{\Omega_\text{L}}
\newcommand{\OmegaII}{\Omega_\text{H}}

\def\bfseries{\fontseries \bfdefault \selectfont \boldmath}

\begin{document}

\newcommand{\titlename}{%
	Model-based {discrete} relaxation process representation of band-limited power-law attenuation}
\title[Discrete relaxations, power-law attenuation]{\titlename}

\author{Sven Peter \surname{N\"asholm}}\affiliation{Department of Informatics, University of Oslo, P.~O.~Box 1080, NO--0316 Oslo, Norway}

\date{November 16, 2012}

\begin{abstract}
Frequency-dependent acoustical loss due to a multitude of physical mechanisms is commonly modeled by multiple relaxations. %
For discrete relaxation distributions, such models correspond with causal wave equations of integer-order temporal derivatives. %
It has {also} been shown that {certain }continuous distributions may give causal wave equations with fractional-order temporal derivatives. %
This paper demonstrates analytically that if the wave-frequency $\omega$ satisfies $\OmegaI\ll \omega \ll \OmegaII$, a continuous relaxation distribution populating only $\Omega\in[\OmegaI,\OmegaII]$ gives the same effective wave equation as for a fully populated distribution. %
This insight sparks {the main contribution:} the elaboration of a method to determine discrete relaxation parameters intended for mimicking a desired attenuation behavior for band-limited waves.  %
In particular, power-law attenuation is discussed as motivated by its prevalence in %
complex media, \eg{} biological tissue. %
A Mittag-Leffler function related distribution {of relaxation mechanisms }has previously been shown to {be related to} the fractional Zener wave equation of three power-law attenuation regimes. %
{Because} these regimes correspond to power-law regimes in the relaxation distribution, %
the idea is to sample the distribution's compressibility contributions evenly in logarithmic frequency while appropriately taking the stepsize into account. %
This work thence claims to provide a model-based approach to determination of discrete relaxation parameters intended to adequately model attenuation power-laws. %
\bigskip

\verb#The peer-reviewed version of this paper is accepted for publishing. It is scheduled#\\
\verb#for Vol. 133, No 3 (March 2013) of The Journal of the Acoustical Society of America.#\\
\verb#DOI: 10.1121/1.4789001 It will be available online at http://asadl.org/jasa/#\\
\verb#  The current document is an e-print which differs in e.g. pagination and typographic#\\
\verb#detail.# 
\end{abstract}
\noindent
							 %
\maketitle


\section{Introduction}

This paper concerns the determination of multiple relaxation parameters. %
%
%

%
%
It is empirically observed that attenuation in biological tissue and other complex media such as polymers, rocks, and rubber often follows a power-law in frequency: $\alpha_k(\omega) \propto \omega^\eta$, with the exponent between 0 and 2 (\cite{Szabo00}). %
Such power-laws can be valid over many frequency decades. %
For acoustic modeling, time-fractional derivative wave equations have been shown to imply power-law attenuation over wide frequency bands (\cite{Holm2010, Holm2011}). %
Such fractional wave equations can be obtained from linearized conservation of mass and momentum in combination with time-fractional constitutive relations connecting stress and strain. %
{Associated} nonlinear fractional wave equations are presented in \cite{Prieur2011}{ and }\cite{Prieur2012}{ while }%
related linear wave-propagation simulations are reported, \eg, in \cite{Wismer1995, Liebler2004, Wismer06, Caputo2011}. %

Moreover, the multiple relaxation mechanism framework of \cite{Nachman1990} is widely considered adequate for acoustic wave modeling in lossy complex media like those encountered in medical ultrasound. %
It relies on thermodynamics and first principles of acoustical physics. %
The corresponding lossy wave equation for $N$ {discrete }relaxation mechanisms is a partial differential equation with its highest time derivative of order $N+2$. %
%
%
%
This model is in the following denoted {the Nachman--Smith--Waag (NSW) model}. %

Viscoelastic constitutive stress-strain models are generally possible to convert into a Maxwell--Wiechert description with springs and dashpots in parallel. %
The NSW model is linked to fractional derivative modeling %
in \cite{Nasholm2011}, where a continuum of relaxation mechanisms is assumed. %
The compressibility contributions were assumed to be distributed following a function related to the Mittag-Leffler function. %
It was shown that the wave equation corresponding to this distribution is identical to the fractional Zener wave equation. %
Actually, a Maxwell--Wiechert description of the fractional Zener stress--strain relation was thereby implicitly verified. %
In \cite{Adolfsson2005}, a very large number of weighted Maxwell elements evenly distributed in the linear frequency scale are shown to give the same stress response as a fractional order viscoelastic model. %
We also note that the rheological fractional spring-pot element was interpreted in terms of weighted springs and dashpots in \cite{Papoulia2010}. %

\cite{Kelly2009} demonstrated that hierarchical fractal ladder networks of springs and dashpots can lead to power-law attenuation in a low-frequency regime. %
This approach however requires a large number of degrees of freedom which makes parameter fits cumbersome. %

%
%
Band-limited fits to power-law acoustic attenuation for relaxation models with $N=2$ and $3$ are exemplified by \cite{Tabei2003} and \cite{Yang2005}. In the latter, one of the mechanisms is assumed to be of very short relaxation frequency, thus representing a thermoviscous component. The two other mechanism relaxation frequencies{, the corresponding two compressibility contributions, and the compressibility of the thermoviscous component are determined through numerical minimization 
of the difference between the resulting attenuation and the desired power-law. }%

For a large number of modeled relaxation mechanisms, such numerical optimization of the parameter fit turns very intricate. %

{Recently published works which stress the need for straightforward and accurate determination of discrete relaxation representations that generate power-law attenuation, are \eg{} }\cite{roitner2012experimental}{ (see Section II) and }\cite{treeby2012modeling}{ (see Section II.C). See also the introduction in }\cite{Liebler2004}{. }%

The present paper establishes a systematic model-based method to choose the compressibility contribution and the relaxation frequency for each of $N$ relaxation processes, to model power-law attenuation over a given wave frequency band. %
%
This {discrete }parameter selection method is based on the continuous distribution of relaxation processes previously shown to result in fractional Zener wave equation, which in turn generates 3 distinctive power-law attenuation regimes (\cite{Nasholm2011}). The present work also studies analytically the effect of letting this distribution cover only a limited frequency band. %

This paper is organized as follows. %
The Theory Section~\ref{sec:Theory} first reviews and partially extends relevant NSW theory, fractional Zener model considerations, and the {link} between those. %
Then it presents{ new developments} related to band-limited continuous and discrete relaxation distributions, as well as the related attenuation for waves of frequencies within and outside the populated relaxation bandwidth.%
{ Section  }\ref{sec:discrete_NSW_bandlim_atten} {  is  central because it describes how to select discrete relaxation process parameters to get approximate power-law attenuation over a given wave-frequency band. }%
Section~\ref{sec:numerical_example} provides two numerical examples exemplifying how discrete relaxation parameters may be determined in order to attain power-law attenuation. Discussions and conclusions are given in Section~\ref{sec:discussion_conslusions}. %

\section{\label{sec:Theory}Theory}

\subsection{Relaxation processes modeling within the NSW framework}
\subsubsection{Discrete relaxation distribution}

The NSW model of multiple discrete relaxation processes results in the frequency-domain generalized compressibility (\cite{Nachman1990})
\begin{align}
	\kappa(\omega) & = \kappaz - i\omega \sum_{\nu=1}^N \dfrac{\kappan \taun}{1 + i\omega\taun}
	= \kappaz - i\omega \sum_{\nu=1}^N \dfrac{\kappan}{\Omega_\nu+ i\omega}
,
	\label{eq:Nachman_kappa_omega}
\end{align}
where the mechanisms $\nu=1\ldots N$, have the relaxation times $\tau_1,\ldots,\tau_N$ and the compressibility contributions $\kappa_1, \ldots, \kappa_N$. In the present work $\Omega_\nu \triangleq 1/\tau_\nu$ denotes the process relaxation frequency. %

The frequency-domain generalized compressibility is in the following simply denoted compressibility. In some branches of science it is {instead }called complex compliance $J^*(\omega)$, being defined as the ratio between strain and stress: $\kappa(\omega) \triangleq \epsilon(\omega)/\sigma(\omega)$. %
It is hence directly related to the constitutive stress--strain relation. %

%

\subsubsection{Continuous relaxation distribution}

Following \cite{Nasholm2011}, a representation of Eq.~\eqref{eq:Nachman_kappa_omega} when considering a continuum of relaxation mechanisms distributed in the frequency band $\Omega \in[\Omega_\text{L},\Omega_\text{H}]$ with the compressibility contributions described by the distribution $\tilde \kappa_\nu(\Omega)$ becomes
\begin{align}
	 \kappa_\text{N}(\omega) \triangleq \kappaz - i\omega\int_{\Omega_\text{L}}^{\Omega_\text{H}} \dfrac{ \tilde\kappa_\nu(\Omega) }{\Omega + i\omega}\, \dd \Omega.
	\label{eq:Nachman_kappa_omega_integral_zenertry}
\end{align}
Letting the integral go from $\Omega_\text{L}=0$ to $\Omega_\text{H}=\infty$, and instead incorporating any possible relaxation distribution bandwidth limitation of $\tilde\kappa_\nu(\Omega)$ into the distribution, we define $\kappa_\nu(\Omega) \triangleq H(\Omega-\Omega_\text{L})H(\Omega_\text{H}-\Omega)\tilde\kappa_\nu(\Omega)$, where $H(\Omega)$ denotes the Heaviside step function. Then the integral \eqref{eq:Nachman_kappa_omega_integral_zenertry} is a Stieltjes transform. {By application of} the Laplace transform relation%
{, which for real $\omega$, real $t$, and real $\Omega$ is valid for $\Omega >0$}: 
%
%
%
%
\begin{align}
	{\mathcal L_t \big\{H(t) e^{-i\omega t}\big\} \!(\Omega) \triangleq \int_0^\infty \! H(t)e^{-i\omega t}\, e^{-\Omega t}\dd t= \frac{1}{ \Omega + i \omega }}
\end{align}
{from the $t$ domain to the $\Omega$ domain, }the compressibility \eqref{eq:Nachman_kappa_omega_integral_zenertry} {may be written as } %
\begin{align}
	\kappa_\text{N}(\omega)
	&= \kappa_0 -i\omega \int_0^\infty \kappan(\Omega) %
		\bigg\{ \int_0^\infty {H(t)} e^{-\Omega t} e^{-i\omega t} \dd t\bigg\}%
	 \dd \Omega,
\end{align}
{which by virtue of Fubini's theorem may be written as }
\begin{align}
	{\kappa_\text{N}(\omega)}
	&{= \kappa_0 -i\omega \int_0^\infty  %
		e^{-i\omega t} \left\{ H(t) \int_0^\infty \kappan(\Omega) e^{-\Omega t} 
		 \dd \Omega\right\} \dd t}\\
	&=  \kappa_0 -i\omega \mathcal F_t \Big\{H(t) \mathcal L_\Omega \left\{ \kappan(\Omega)  \right\}\!\! (t) \Big\}(\omega),
	\label{eq:nachman_fourier_laplace}
\end{align}
{where the Fourier transform from the $t$ domain to the $\omega$ domain is $\displaystyle \mathcal F_t\left\{ f(t) \right\}(\omega) \triangleq \int_{-\infty}^\infty e^{-i\omega t} f(t)\; \dd t$. }

Another {equivalent }representation of the compressibility \eqref{eq:Nachman_kappa_omega_integral_zenertry} is
\begin{align}
	\kappa_\text{N}(\omega) &= \kappa_0-\omega^2\int_{0}^{\infty} \dfrac{\kappan(\Omega)}{\Omega^2+\omega^2}\dd \Omega -i\omega  \int_{0}^{\infty} \dfrac{\Omega\kappan(\Omega)}{\Omega^2+\omega^2} \dd \Omega,
	\label{eq:nachman_expanded}
\end{align}
which is found by multiplying both the numerator and the denominator of Eq.~\eqref{eq:Nachman_kappa_omega_integral_zenertry}{ by $(\Omega - i\omega)$. }

\subsubsection{Attenuation and phase velocity}

The conventional decomposition of the frequency-dependent wavenumber $k(\omega)$ into its real and imaginary parts, 
gives the phase velocity $c_p(\omega) = \omega/\Re\left\{k(\omega)\right\}$ and the attenuation $\alpha_k(\omega) = -\Im\left\{k(\omega)\right\}$. %

Combining the definition of $\kappa(\omega)\triangleq \epsilon(\omega)/\sigma(\omega)$ with the linearized conservation of mass and momentum (see \cite{Nasholm2011} and the references therein for details) gives 
\begin{align}
	k^2&(\omega) = \omega^2\rho_0 \kappa(\omega). 
	\label{eq:dispersion_relation}
\end{align}
In general, the attenuation and the phase velocity are thus given from the dispersion relation above as
\begin{align}
\begin{array}{l}
	\alpha_k(\omega) = -\Im\left\{k\right\} = -\omega\sqrt{\rho_0}\Im\left\{\sqrt{\kappa(\omega)}\right\}	\quad\text{and}\\
	c_p(\omega) = \omega/\Re\left\{k\right\} = {\rho_0^{-1/2}}/\Re\left\{\sqrt{\kappa(\omega)}\right\}.
	\label{eq:atten_and_soundspeed_from_kappa}
	\end{array}
\end{align}

{Analyzing this expression further, we see that i}n the small-attenuation regime where the {$\kappa_0$} part of \eqref{eq:nachman_expanded} dominates over the imaginary part {infers}
\begin{align}
	&{\sqrt{\kappa_\text{N}(\omega)} }\notag\\
	 & {=\sqrt{\kappa_0-\omega^2\int_{0}^{\infty} \dfrac{\kappan(\Omega)}{\Omega^2+\omega^2}\dd \Omega -i\omega  \int_{0}^{\infty} \dfrac{\Omega\kappan(\Omega)}{\Omega^2+\omega^2} \dd \Omega}}\notag\\
	 &{\approx \sqrt{\kappa_0} -\dfrac{1}{2\sqrt{\kappa_0}}\left[ -\omega^2\int_{0}^{\infty} \dfrac{\kappan(\Omega)}{\Omega^2+\omega^2}\dd \Omega -i\omega  \int_{0}^{\infty} \dfrac{\Omega\kappan(\Omega)}{\Omega^2+\omega^2} \dd \Omega\right]} 
\end{align}
{Then the attenuation }$\alpha_k$ is given from the approximation
\begin{align}
{\alpha_k(\omega) = }
	&-\omega\sqrt{\rho{_0}}\Im\left\{ \sqrt{\kappa_\text{N}(\omega)} \right\}  \notag\\%
	&\approx {A}\omega^2  \int_{0}^{\infty} \dfrac{\Omega\kappan(\Omega)}{\Omega^2+\omega^2} \dd \Omega \label{eq:alpha_k_smallatten}\\ 
	& = {A}\omega^2 \int_{0}^\infty \int_0^\infty \kappa(\Omega)  e^{-\Omega t} \cos(\omega t)\;\dd t\;\dd \Omega\notag\\
	& = {A}\omega^2 \mathcal{F}_\text{c} \big\{\mathcal{L}{_\Omega}\big\{ \kappa_\nu(\Omega) \big\}(t) \big\}(\omega),
	\label{eq:nachman_expanded_smallatten}
\end{align}
where $\mathcal F_\text{c}$ denotes the Fourier cosine transform {from the $t$ domain to the $\omega$ domain, }and $\mathcal L{_\Omega}$ {denotes }the Laplace transform{ from the $\Omega$ domain to the $t$ domain}. %
{The introduced $A$ is a frequency-independent scalar. }%
The integral of Eq.~\eqref{eq:alpha_k_smallatten}{ is a Widder potential transform (}\cite{widder1966transform}{). }
%
%
%
 
Using Eq.~\eqref{eq:atten_and_soundspeed_from_kappa} and \eqref{eq:nachman_expanded_smallatten} under the small-attenuation constraint, the relaxation process spectrum $\widetilde \kappa_\nu(\Omega)$ corresponding to some frequency-dependent attenuation model $\widetilde \alpha_k(\omega)$ may {hence }be constructed using the {inverse transform }recipe
\begin{align}
	\widetilde \kappa_\nu(\Omega) = A \cdot \mathcal{L}^{-1}_t \left\{     \mathcal{F}_\text{c}^{-1} \left\{ \dfrac{\widetilde\alpha_k(\omega) }{\omega^2}\right\}(t)\right\}(\Omega){.}
	\label{eq:kappa_from_alpha}
\end{align}
{E}xpression \eqref{eq:kappa_from_alpha} is similar to {what was found using }an approach reported in \cite{Vilensky2012}, %
however a formula equivalent to \eqref{eq:alpha_k_smallatten} was used already in \eg{} \cite{Pauly1971}. %
The small-attenuation assumption is probably reasonable for compressional wave propagation in biological tissue. By contrast, for shear-wave propagation, the attenuation is generally much more pronounced (\cite{Szabo00}). %

For a discrete set of relaxation mechanisms under the small-approximation assumption, %
we see from Eq.~\eqref{eq:alpha_k_smallatten} that the total attenuation is just the sum of the contribution from each mechanism: $\alpha_k(\omega) = \sum_{\nu=1}^N \alpha_\nu(\omega)$, where the contribution of mechanism $\nu$ is given by
\begin{align}
	\alpha_\nu(\omega) = \dfrac{\kappa_\nu\Omega_\nu \omega^2}{\omega^2 +\Omega_\nu^2}.
	\label{eq:alpha_k_NSW_decoupled}
\end{align}
Such behavior is especially well documented in air (\cite{Bass1995}) and in sea water (\cite{Ainslie1998}). %

\subsection{The fractional Zener wave equation}
%
%
As a consequence of the fractional Zener model stress--strain relation (see \eg{} \cite{Bagley83A}), the frequency-domain fractional Zener compressibility is obtained from the ratio $\epsilon(\omega)/\sigma(\omega)$:
\begin{align}
	\kappa_\text{Z}(\omega) & \triangleq %
		\kappaz \frac{1 + (\tau_{\epsilon}i \omega)^{\beta}}{1+ (\tau_{\sigma}i \omega)^{\alpha}}. %
	\label{eq:fZener_compressibility}
\end{align}
Due to thermodynamic constraints, $0<\alpha\leq 1$ and $\beta\leq\alpha$ (\cite{Glockle1991}){, however the case $\alpha=\beta$ is the most well-behaved from a physical point of view }\cite{rossikhin2001analysis}{. }%
Insertion of the compressibility \eqref{eq:fZener_compressibility} into the dispersion relation \eqref{eq:dispersion_relation}, generates the fractional Zener dispersion relation (\cite{Holm2011})
\begin{align}
	k^2 %
	= \frac{\omega^2}{c^2_0}\frac{1 + (\tau_{\epsilon}i \omega)^{\beta}}{1+ (\tau_{\sigma}i \omega)^{\alpha}}.
	\label{Eq:k2}
\end{align}
This is the spatio-temporal frequency domain representation of the five-parameter fractional Zener wave equation which by inverse transform hence becomes
\begin{align}
	{\nabla^2 u -\dfrac 1{c_0^2}\frac{\partial^2 u}{\partial t^2} + \taus^\alpha \dfrac{\partial^\alpha}{\partial t^\alpha}\nabla^2 u	- \dfrac {\taue^\beta}{c_0^2} \dfrac{\partial^{\beta+2} u}{\partial t^{\beta+2}} = 0.}
	\label{Eq:wave_equation_zener}
\end{align}
%
%
%
%

%
The $\alpha=\beta$ variant of the fractional Zener compressibility \eqref{eq:fZener_compressibility} combined with Eq.~\eqref{eq:atten_and_soundspeed_from_kappa} results in three distinct attenuation power-laws (\cite{Holm2011}): 
\begin{align}
	\alpha_k \propto %
	\left\{
	\begin{array}{ll}
		\omega^{1+\alpha} & \text{low-frequency regime,} \\
		\omega^{1-\alpha/2} & \text{intermediate frequency regime,}\\
		\omega^{1-\alpha} & \text{high-frequency regime.}
	\end{array}
	\right.
	\label{eq:freq_regions}
\end{align}

%
%

%
%
%

\subsection{Connecting the NSW and the fractional Zener models\label{sec:connectin_NSW_Zener}}
%
%
\subsubsection{The continuous relaxation distribution $\kappa_{\nu}'(\Omega)$\label{sec:xxxxxxxxxx}}
Provided that the linearized conservations of mass and momentum are valid, and given that the NSW compressibility $\kappa_\text{N}(\omega)$ of Eq.~\eqref{eq:nachman_fourier_laplace} 
is equal to the fractional Zener compressibility $\kappa_\text{Z}(\omega)$ of Eq.~\eqref{eq:fZener_compressibility}, %
the dispersion relations from Eq.~\eqref{eq:dispersion_relation} are also equal. Because the dispersion relation is a spatio-temporal Fourier representation of the wave equation, %
$\kappa_\text{N}(\omega) = \kappa_\text{Z}(\omega)$ thus implies that the NSW wave equation becomes equal to the fractional Zener wave equation \eqref{Eq:wave_equation_zener}. %
As brought forward in \cite{Nasholm2011}, there exists a continuous distribution $\kappa_\nu'(\Omega)$ of NSW relaxation processes which implies $\kappa_\text{N}(\omega) = \kappa_\text{Z}(\omega)$ when $\alpha=\beta$:
\begin{align}
	\kappa_\nu'(\Omega) %
	& \triangleq \dfrac{1}{\pi} \dfrac{\kappaz(\taus^{\alpha}- \taue^\alpha)\Omega^{\alpha-1} \sin (\alpha\pi ) }{ (\taus\Omega)^{2\alpha} + 2(\taus\Omega)^\alpha \cos(\alpha\pi) + 1}
	.
	\label{eq:distribution_for_alphaisbeta}
\end{align}
This distribution is the inverse Laplace transform of a Mittag-Leffler related expression, see \cite{Nasholm2011} for details. %
Note that this link between the fractional Zener and the NSW models is valid also outside the small-attenuation regime $\Im\left\{k\right\} \ll \Re\left\{k\right\}$. %

\subsubsection{$\kappa_{\nu}'(\Omega)$ power-law regimes\label{sec:power-law_regimes}}
%
%
Below follows an analysis of the relaxation time distribution $\kappa_\nu'(\Omega)$ of Eq.~\eqref{eq:distribution_for_alphaisbeta} as it evolves depending on $\Omega \taus$:
\begin{align}
	\kappa_{\nu}'(\Omega) %
	 &= \dfrac{1}{\pi} \dfrac{\kappa_0(\tau_\sigma^{\alpha}- \tau_\epsilon^\alpha)\Omega^{\alpha-1} \sin (\alpha\pi ) }{ (\tau_\sigma\Omega)^{2\alpha} + 2(\tau_\sigma\Omega)^\alpha \cos(\alpha\pi) + 1}\notag\\
	 & \approx \left\{
		\begin{array}{ll}
			\displaystyle C_\text{L}  \cdot\  \Omega^{\alpha-1}, & \ \text{for } \Omega\taus \ll 1\\
			\displaystyle C_\text{I} \cdot\ \Omega^{-1}, & \ \text{for } \Omega\taus \approx 1\\
			\displaystyle C_\text{H} \cdot\ \Omega^{-\alpha-1},   & \  \text{for } 1 \ll \Omega\taus,
		\end{array}
		\right. 
	\label{eq:chosen_kappa_n_Omega}
\end{align}
where the frequency-independent constants are:%
\begin{align}
	\begin{array}{ll}
	C_\text{L} & \triangleq \dfrac{\kappa_0(\tau_\sigma^{\alpha}- \tau_\epsilon^\alpha)\sin(\alpha\pi) }{ \pi},\\
	C_\text{I} & \triangleq \dfrac{\kappa_0(\tau_\sigma^{\alpha}- \tau_\epsilon^\alpha)\sin(\alpha\pi) }{2 \pi\tau_\sigma^{\alpha}(1+\cos(\alpha\pi)) }, \text{ and}\\
	C_\text{H} & \triangleq \dfrac{\kappa_0(\tau_\sigma^{\alpha}- \tau_\epsilon^\alpha)\sin (\alpha\pi ) }{\pi\tau_\sigma^{2\alpha}}.
	\end{array}
\end{align}

Due to the restriction $0<\alpha \leq 1$, we hence note that depending on the value of the product $\taus\Omega$, the distribution $\kappa_\nu'(\Omega)$ may have the form of a power-law $\kappa_\nu' \propto \Omega^d$, where $-2<d\leq 0$. %

\subsection{A band-limited continuum of NSW relaxation processes\label{sec:band-limited_continuum}}

In the following, the case of a continuum of relaxation processes populating a limited frequency band $\Omega \in \left[ \Omega_\text{L},\Omega_\text{H}\right ]$ is further explored {with respect to the  for three different wave-frequency regimes}. %

\subsubsection{Formal analysis of band-limited power-law relaxation distribution}
The case of a relaxation process continuum populating the region $\Omega\in[\Omega_\text{L},\Omega_\text{H}]$, for example where $\kappa_\nu(\Omega)$ is adequately be approximated by a power-law (see Section~\ref{sec:power-law_regimes} above)%
\begin{align}
	\kappa_\nu(\Omega) = \left\{
		\begin{array}{ll}
			C\cdot\Omega^d  &\qquad\Omega\in[\Omega_\text{L},\Omega_\text{H}],\\
			0 & \qquad\text{otherwise.}
		\end{array}
		\right.
\end{align}
{where $-2<d\leq 0$,} is now investigated. %
Distributions of such form are related to broken power-laws, often denoted Pareto distributions{, }and are commonly observed in nature and society (\cite{newman2005}). %

For $\kappa_\nu(\Omega) = C\cdot \Omega^d$, {the generalized compressibility }\eqref{eq:Nachman_kappa_omega_integral_zenertry} may be written
\begin{align}
	 \kappa_\text{N}(\omega) & = \kappaz - i\omega C \!\!\left[\int_{\Omega_\text{L}}^{\infty} \dfrac{ \Omega^d }{\Omega + i\omega}\, \dd \Omega
	- \int_{\Omega_\text{H}}^{\infty} \dfrac{ \Omega^d }{\Omega + i\omega}\, \dd \Omega\right],
	\label{eq:broken_powerlaw}
\end{align}
which in view of Eq.~\eqref{eq:nachman_expanded} and \cite{bateman1954V1C6} [Eq.~(21)], becomes
\begin{align}
	\kappa&_\text{N}(\omega)  = \kappaz + \dfrac{C\omega^2}{2}\left[ \Omega_\text{H}^{d-1} \Lerch{-\dfrac{\omega^2}{\Omega_\text{H}^2}}{1}{\dfrac{1-d}{2}}
	\right.\notag\\
	&\left.  -\Omega_\text{L}^{d-1} \Lerch{-\dfrac{\omega^2}{\Omega_\text{L}^2}}{1}{\dfrac{1-d}{2}}
	\right]\notag\\
	&+ \dfrac{iC\omega}{2}\left[ \Omega_\text{H}^{d} \Lerch{-\dfrac{\omega^2}{\Omega_\text{H}^2}}{1}{-\dfrac{d}{2}}
	 -\Omega_\text{L}^{d} \Lerch{-\dfrac{\omega^2}{\Omega_\text{L}^2}}{1}{-\dfrac{d}{2}}
	\right],
	\label{eq:lercrelated}
\end{align}
where $\Lerch{x}{\nu}{u}$ is the analytic continuation of the Lerch transcendent. %
{In Section }\ref{sec:omega_inbetween}{ below, the compressibility }\eqref{eq:lercrelated}{ is further analyzed. First a continuous band-limited power-law relaxation spectrum is considered and the corresponding generalized compressibility is deduced. Then three different wave-frequency regimes are considered for any band-limited relaxation relaxation $\kappa_\nu(\Omega)$ and the corresponding attenuation functions are brought forth. }
For $\nu=1$, {the function $\Lerch{x}{\nu}{u}$} is related to the analytic continuation of the Gauss hypergeometric function as: (\cite{bateman1954V1C6}, Eq.~(1.10))
\begin{align}
	\Lerch{x}{1}{u} = u^{-1}{}_2F_1(1,u; 1+u; x),
\end{align}
to which there is a direct connection to fractional calculus because (\cite{samko1993chapter2})
\begin{align}
	{}_2F_1(a,b; c; x) = \dfrac{\Gamma(a) x^{1-c}}{\Gamma(b)} {}_0 D_x^{1-c} \dfrac{x^{b-1}}{(1-x)^a}.
\end{align}

For practical purposes when $\omega$ is far from the relaxation process cut-off frequencies $\Omega_\text{L}$ and $\Omega_\text{H}$, it is however not necessary to evaluate the Lerch transcendents of \eqref{eq:lercrelated} to find $\kappa(\omega)$ and hence determine the frequency-dependent attenuation and phase velocity. The three relevant frequency regimes are considered in the following two subsections.  

\subsubsection{The $\Omega_\text{L} \ll \omega \ll \Omega_\text{H}$ frequency range for any $\kappa_\nu(\omega)$\label{sec:omega_inbetween}}
In case the wave-frequency is within the frequency band of the relaxation processes, in the integrals of \eqref{eq:nachman_expanded} we change the integration limits from $0$ into $\Omega_\text{L}$ and from $\infty$ into $\Omega_\text{H}$ and in addition make the variable change $u\triangleq \Omega/\omega$ to result in
\begin{align}
	\kappa_\text{N}(\omega) &= \kappa_0-\omega\int_{\frac{\scriptstyle \Omega_\text{L}}{\scriptstyle \omega}}^{\frac{\scriptstyle \Omega_\text{H}}{\scriptstyle \omega}} \dfrac{\kappan(u\omega)}{u^2+ 1}\dd u -i\omega  \int_{\frac{\scriptstyle \Omega_\text{L}}{\scriptstyle \omega}}^{\frac{\scriptstyle \Omega_\text{H}}{\scriptstyle \omega}} \dfrac{u\kappan(u\omega)}{u^2+ 1} \dd u.  
	\label{eq:nachman_expanded_approx}
\end{align}
For the analyzed case $\Omega_\text{L} \ll \omega \ll \Omega_\text{H}$, the integration limits then become $\Omega_\text{L}/\omega \rightarrow 0$ and $\Omega_\text{H}/\omega \rightarrow \infty$. %
Then performing the variable change back into $\Omega=u\omega$ makes it clear that under the given conditions the effective compressibility thus becomes equal to the compressibility related to the fully-populated relaxation frequency band \eqref{eq:nachman_expanded}. %
A similar analysis may also be done directly in \eqref{eq:Nachman_kappa_omega_integral_zenertry}. %
The fractional Zener wave equation \eqref{Eq:wave_equation_zener} and its associated attenuation and phase velocity expressions %
(see \cite{Nasholm2011}) %
are consequently valid also for the band-limited distribution of relaxation processes, as long as the wave frequency $\omega$ is much lower than $\Omega_\text{H}$ and much greater than $\Omega_\text{L}$. %

{Interestingly, the analysis above prompts the conclusion that for a band-limited %
($\Omega_\text{L} \ll \omega \ll \Omega_\text{H}$) %
power-law relaxation distribution $\kappa_\nu(\Omega) \propto \Omega^d$, which results in the compressibility $\kappa_\text N(\omega)$ given by %
}
\eqref{eq:lercrelated}{, there is a direct relation to the attenuation power-law valid within $\Omega_\text{L} \ll \omega \ll \Omega_\text{H}$: }%

{First, consider the low-frequency power-law regime of the distribution $\kappa_\nu'(\Omega)$ in Eq.}~\eqref{eq:distribution_for_alphaisbeta} %
{being populated in the limited relaxation frequency band. }%
Then $\kappa_\nu'(\Omega) \propto \Omega^{d}$ which gives $d=(\alpha-1) \in [-1,0]$, because $\alpha \in[0,1]$. By virtue of Eq.~\eqref{eq:freq_regions}%
{ this corresponds to the attenuation power-law $\alpha_k(\omega) \propto \omega^\eta$ with the exponent $\eta \in [1+\alpha] \Leftrightarrow \eta \in[1,2]$. }%
{Second, consider instead the high-frequency regime of the distribution $\kappa_\nu'(\Omega)$ being populated in the limited relaxation frequency band. This leads to $\kappa_\nu'(\Omega) \propto \Omega^d$ with $d = (-\alpha-1) \in[-2,-1]$. }%

{Hence a band $\Omega_\text L \ll \omega \ll \Omega_\text H$ populated with relaxation frequency contributions $\kappa_\nu(\Omega) \propto \Omega^d$ gives a band-limited power-law attenuation $\alpha_k(\omega) \propto \omega^{d+2}$ both for $d\in[-2,-1]$ and for $d\in[-1,0]$. }

\subsubsection{The $\omega \ll \Omega_\text{L} < \Omega_\text{H}$ and $\Omega_\text{L} < \Omega_\text{H} \ll \omega$ frequency regimes for any $\kappa_\nu(\omega)$}
As pointed out in \cite{Nasholm2011}, the case of the wave-frequency being much lower than the populated relaxation process frequency band $\Omega\in[\Omega_\text{L},\Omega_\text{H}]$, an analysis similar to the LF limit of \cite{Nachman1990} results in $\alpha_k \propto \omega^2$. By the same token, for the wave-frequency being much higher than the populated relaxation proportional to frequency band, the attenuation is frequency-independent. %

Within these frequency regimes, the effective wave equation is hence the same as for a single discrete relaxation mechanism with the compressibility contribution being proportional to the impulse $\delta(\Omega-(\Omega_\text{L}+\Omega_\text{H})/2)$. This is similar as for a medium with one single discrete NSW relaxation process. %

\subsection{$N$ discrete NSW relaxation processes to get band-limited attenuation power-law\label{sec:discrete_NSW_bandlim_atten}}
Based on the developments in Section~\ref{sec:band-limited_continuum}, here a method is {elaborated } to determine the $\Omega_\nu$ and $\kappa_\nu$ parameters of a discrete set of $N$ relaxation processes intended to result in an appropriate attenuation power-law $\alpha_k(\omega) \propto \omega^{\eta}$ for waves within the bandwidth $\omega\in[\Omega_\text{L},\Omega_\text{H}]$. The power-law exponent satisfies $\eta \in[0,2]$. %

First, based on \eqref{eq:freq_regions}, the $\eta$ exponent determines whether a low, high, or intermediate frequency model is applied. %
Based on this, $\taus$ is then set so that $1/{\taus}$ is either much higher or much lower than the frequency region of interest $[\Omega_\text{L}, \Omega_\text{H}]$. %
Subsequently, the relaxation process frequencies $\Omega_\nu$ are sampled within $[\Omega_\text{L}, \Omega_\text{H}]$, equi-spaced in the $\log \Omega$ domain hence giving %
\begin{align}
	\Omega_\nu =  \Omega_\text{L}^{\frac{N-\nu}{N-1}} \Omega_\text{H}^{\frac{\nu-1}{N-1}}.
\end{align} 
Thereafter, the compressibility contribution $\kappa_\nu(\Omega_\nu)$ of each process is decided from $\kappa_\nu'(\Omega_\nu)$ of \eqref{eq:distribution_for_alphaisbeta}. %
Alternatively, the relevant power-law approximate expression of \eqref{eq:chosen_kappa_n_Omega} may be chosen. %
Finally, the $\taue$ parameter is adjusted to achieve the attenuation $\alpha_k(\omega_\text{ref}) = \alpha_\text{ref}$ at some appropriate reference frequency $\omega_\text{ref} \in [\Omega_\text{L}, \Omega_\text{H}]$. %

For calculation of the resulting compressibility $\kappa_\text{N}(\omega)$, the discretized approximation of the integral \eqref{eq:Nachman_kappa_omega_integral_zenertry} must take into account the uneven stepsize 
\begin{align}
	\Delta\Omega_\nu = \Omega_\nu\left(1-\left(\Omega_\text{L}/\Omega_\text{H}\right)^{1/(N-1)}\right).
\end{align}
The $\kappa_\text{N}(\omega)$ estimate $\hat \kappa_\text{N}(\omega)$ which approximates \eqref{eq:Nachman_kappa_omega_integral_zenertry} for the chosen discrete relaxation process parameters is hence
\begin{align}
	 \hat\kappa_\text{N}(\omega) = \kappaz - i\omega\sum_{\nu=1}^N \dfrac{\kappa_\nu(\Omega_\nu) }{\Omega_\nu + i\omega} \Delta\Omega_\nu.
	\label{eq:Nachman_kappa_omega_discretized_unevenly}
\end{align}
From this compressibility, the attenuation and phase velocity may be calculated in the conventional manner using Eq.~\eqref{eq:atten_and_soundspeed_from_kappa}. %
The attenuation resulting from this relaxation process parameter decision approach is explored in the following 2 numerical examples.

\section{\label{sec:numerical_example}Numerical examples}
%
%
%
%
\begin{table}[t]
	\caption{\label{tab:parameters}Medium parameters for the attenuation power-law fit of Section~\ref{sec:Cleveland_attenuation}, similar to the \cite{Yang2005} parameters.}
	\begin{ruledtabular}
	\begin{tabular}{l r l c}
		Equilibrium speed of sound, $c_0$ (m/s) & 1540 \\
		Density, $\rho_0$ (kg/m$^3)$ & 1050 \\
		Zero-freq.~compressibility,  $\kappaz=\frac 1 {c_0^2\rho_0}$ (Pa$^{-1})$  & \hspace{-.7cm}$4.0158\cdot 10^{-10}$ \\
		Wanted attenuation at 1 MHz, $\alpha_0$  (dB/MHz/cm)  & 0.3 \\
		Wanted attenuation power-law exponent, $\eta$ & 1.1\\
	\end{tabular}
	\end{ruledtabular}
\end{table}

\subsection{Power-law $\alpha_k \propto \omega^{1.1}$ for $f\in[100\;\text{kHz}, 30\;\text{MHz}]$\label{sec:Cleveland_attenuation}}

\begin{figure}[!t]
	\begin{center}
		\insfig{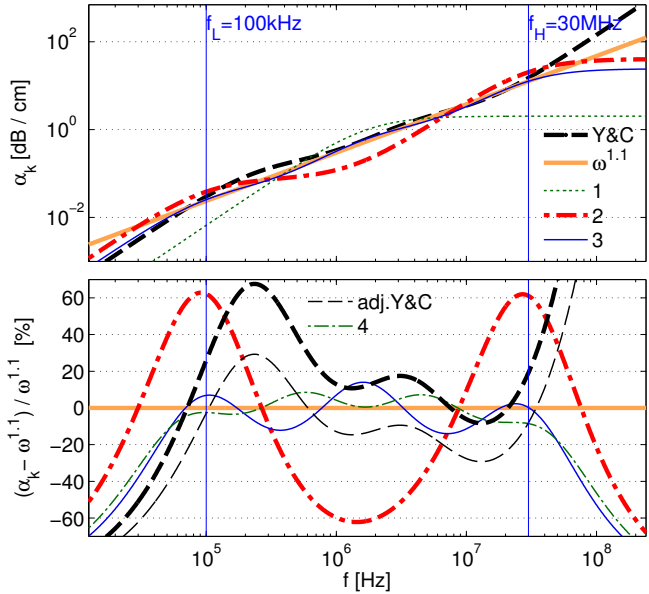}
	\end{center}
\caption{\label{fig:attenuation_model_parameter}%
(Color online) %
Attenuation as a function of frequency for $N$ o-relaxation mechanisms determined by the method proposed in Section~\ref{sec:discrete_NSW_bandlim_atten} aiming at constructing a power-law valid within 100\:kHz and 30\:MHz. %
Also displayed are the \cite{Yang2005} $N=3$ attenuation and an ideal power-law $\propto \omega^{1.1}$ (thick solid line). %
Table~\ref{tab:parameters} lists the medium properties. %
Top pane: Attained attenuations. %
Bottom pane: Relative difference between the attained relaxation attenuation functions and the power-law. %
The horizontal axes represent wave-frequency. For visualization convenience, each attenuation is normalized in order to make the minimum and maximum relative differences equal in magnitude within the relevant frequency interval. The Yang and Cleveland attenuation is included both with normalization (thick dashed line) and without (thin dashed line, only included in the bottom pane). %
The attenuation is shown for $N=1$ (thin dashed line), $N=2$ (thick dash-dotted line), and $N=3$ (thin solid line). 
The relaxation parameters $\Omega_\nu$ and $\kappa_\nu$ are displayed in Table~\ref{tab:omega_nu_and_kappa_nu}.
} 
\end{figure}

Below follows an explicit example of modeling an attenuation power-law $\alpha_k \propto \omega^{1.1}$ between 100\:kHz and 30\:MHz using discrete relaxation processes. Such attenuation is relevant for medical ultrasound imaging. The chosen medium properties, which are listed in Table~\ref{tab:parameters}, are the same as in \cite{Yang2005}. %
The attenuation resulting from the parameter selection approach proposed in Section~\ref{sec:discrete_NSW_bandlim_atten}, was applied for $N=1,\ldots,4$ relaxation mechanisms using $\Omega_\text{L}=100$\;kHz and $\Omega_\text{H}=10$\;MHz. %
The consequent attenuation functions as calculated from the {NSW} compressibility in Eq.~\eqref{eq:Nachman_kappa_omega}%
{ }are displayed in Fig.~\ref{fig:attenuation_model_parameter}, where also a pure $\alpha_k(\omega) \propto \omega^{1.1}$ law is plotted, as well as the Yang and Cleveland attenuation. %
In \cite{Yang2005}, a numerical least squares scheme was used to fit two relaxation terms and a thermoviscous component to an approximate $\alpha_k\propto \omega^\eta$ power-law using Eq.~\eqref{eq:alpha_k_NSW_decoupled}. The relaxation frequency $\Omega_\nu$ was pre-set for the thermoviscous component, therefore leaving two $\Omega_\nu$ and three $\kappa_\nu$ parameters to be determined by the numerical scheme. %
The {discrete }relaxation frequencies and compressibility contributions are displayed in Table~\ref{tab:omega_nu_and_kappa_nu}. %
%
\begin{table}[t!]
	\caption{\label{tab:omega_nu_and_kappa_nu}Relaxation process frequencies $\Omega_\nu$ and compressibility contributions $\kappa_\nu$ corresponding to the attenuation functions displayed in Fig~\ref{fig:attenuation_model_parameter}. The parameters were attained using the Section~\ref{sec:discrete_NSW_bandlim_atten} method for $N=1,2,3$ relaxation mechanisms. }
	\begin{ruledtabular}
	\begin{tabular}{l | c c c |c  c c c}
		& $\Omega_1$ & $\Omega_2$ &$\Omega_3$ & $\kappa_1$ &$\kappa_2$ &$\kappa_3$ \\
		& (MHz) &(MHz) &(MHz) & (TPa)$^{-1}$& (TPa)$^{-1}$& (TPa)$^{-1}$\\
		\hline
		$N=1$ & 1.7& & & 2.6\\
		$N=2$ & 0.10& 3.0  &     & 1.7 & 3.1 \\
		$N=3$ & 0.10& 1.73 & 3.0 & 1.0 & 1.3 & 1.8
	\end{tabular}
	\end{ruledtabular}
\end{table}

{A fit for the same attenuation law to the fractional Zener model and to the corresponding continuous compressibility function is provided in }\cite{Nasholm2011}{ where Table II lists the parameters $\alpha$, $\tau_\epsilon$, $\tau_\sigma$, and $\kappa_0$. }%

\subsection{Power-law $\alpha_k \propto \omega^{1.1}$ for $f\in[100\:\text{kHz}, 1\:\text{GHz}]$}

\begin{figure}[!t]
	\begin{center}
		\insfig{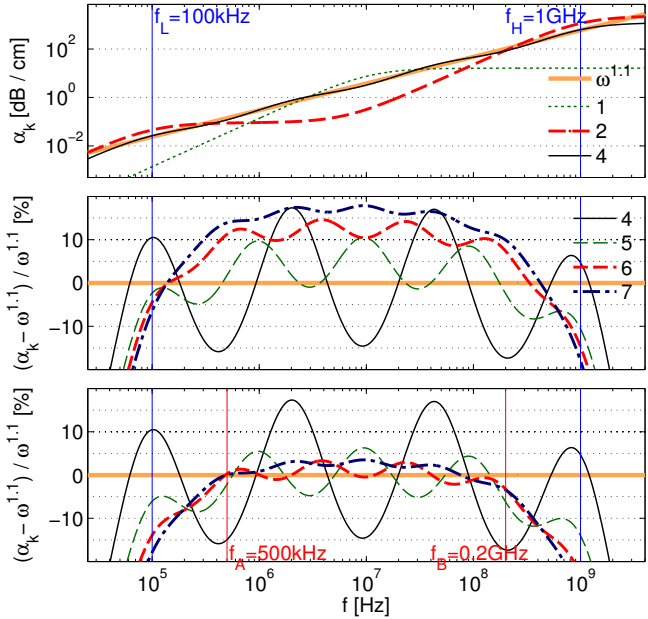}
	\end{center}
\caption{\label{fig:attenuation_model_wid}(Color online) %
Attenuation as a function of frequency for $N$ relaxation mechanisms determined by the method proposed in Section~\ref{sec:discrete_NSW_bandlim_atten} aiming at constructing a power-law $\propto \omega^{1.1}$ (thick solid line) valid within the wide interval between 100\:kHz and 1\:GHz. %
The horizontal axes represent wave-frequency. %
Top pane: the attenuation functions for $N=1$ (thin dashed line), $N=2$ (thick dashed line), and $N=4$ (thin solid line). %
Middle pane: Relative difference between the attained relaxation attenuation functions and the power-law, with each attenuation normalized in order to make the minimum and maximum relative differences equal in magnitude within 100\:kHz and 1\:GHz for $N=4$ (thin solid line), $N=5$ (thin dashed line), $N=6$ (thick dashed line), and $N=7$ (thick dash-dotted line). %
Bottom pane: Same as the middle pane, however with the normalization taking the more narrow frequency interval between $500$\:kHz and $0.2$\:GHz into account. %
} 
\end{figure}

In the following, the approach suggested in Section~\ref{sec:discrete_NSW_bandlim_atten} is applied in a similar way as in Section~\ref{sec:Cleveland_attenuation} to construct power-law attenuation within the wider wave-frequency band $f\in[100\:\text{kHz}, 1\:\text{GHz}]$. This frequency band covers more or less all frequencies of conventional pulse-echo medical ultrasound imaging and microscopy. %
The properties of discrete relaxations with $\Omega_\nu$ evenly distributed in $\log \Omega$ were determined for distribution sets %
with 1 to 7 mechanisms. %
The attenuation functions corresponding to each relaxation mechanism set are displayed in Fig.~\ref{fig:attenuation_model_wid}. %

\section{\label{sec:discussion_conslusions}Discussion and Concluding remarks}

As laid out in Section~\ref{sec:band-limited_continuum}, the present paper demonstrates that for a continuous distribution of relaxation processes with the compressibility contribution distribution $\kappa_\nu(\Omega)$ populating the relaxation frequency range $\Omega\in[\Omega_\text{L}, \Omega_\text{H}]$, %
the effective wave equation for the wave-frequency $\omega$ satisfying $\OmegaI \ll \omega \ll \OmegaII$ is the same as for the relaxation processes populating the whole $\Omega\in[0,\infty]$. %
This work thus supports the intuitive conjecture that for a distribution of relaxation processes covering all frequencies, it is the $\Omega_\nu$ within the wave-frequency bandwidth that mainly contribute to the attenuation. The wave may thus be seen as probing the medium around the relaxation frequencies within and close to the wave-frequency bandwidth. %
 
When instead $\omega \ll \OmegaI<\OmegaII$ or $\OmegaI<\OmegaII\ll \omega$, %
the effective wave equation becomes the same as for a single discrete NSW relaxation mechanism at $\Omega_\nu = (\OmegaI+\OmegaII)/2$. %
All attenuation models that may be written as a superposition of NSW relaxations are causal. %

In particular this work considers the Mittag-Leffler function related distribution case $\kappa_\nu'(\Omega)$ as given in \eqref{eq:chosen_kappa_n_Omega}. For a fully populated range $\Omega\in[0,\infty]$ this has previously been shown to produce the four-parameter fractional Zener wave equation prompting three distinct frequency power-law attenuation regimes. %
If this relaxation mechanism distribution instead populates only $\Omega\in[\OmegaI,\OmegaII]$, {as shown in Section }\ref{sec:omega_inbetween}, the 4-parameter fractional Zener wave equation is still valid as long as the wave-frequency $\omega$ satisfies $\OmegaI \ll \omega \ll \OmegaII$. %
Actually, the distribution $\kappa_\nu'(\Omega)$ may, as shown in Section~\ref{sec:power-law_regimes}, for three different $\Omega$ regions be appropriately approximated by $\propto \omega^d$, with $-2<d\leq 0$. %
These regions are linked to the attenuation power-law frequency regimes of the 4-parameter fractional Zener wave equation. %

{The insights of Sections~}\ref{sec:connectin_NSW_Zener}{ and }\ref{sec:band-limited_continuum}{ }provide{ }the basis upon which a proposed method for selection of discrete relaxation distribution properties relies. %
{The discrete relaxation parameter parameter selection approach is laid out in Section }\ref{sec:discrete_NSW_bandlim_atten}{, which is intended to be a key advance of the current paper. }
One may hence avoid applying numerical minimization methods to determine the relaxation parameters, which becomes cumbersome especially when there is a large number of mechanisms. %
This proposed selection of discrete parameters is {model}-based primarily in the sense that for a chosen process relaxation frequency $\Omega_\nu$, the compressibility $\kappa_\nu$ is decided from the distribution $\kappa_\nu'(\Omega_\nu)$ given in \eqref{eq:chosen_kappa_n_Omega}. %
The $2N$-parameter determination task to decide the $\Omega_\nu$ and $\kappa_\nu$ values for a model of $N$ relaxation mechanisms thus becomes an $N$-parameter determination problem. 
However for determination of the $N$ relaxation frequency parameters, the straightforward choice to distribute  $\Omega_\nu$ logarithmically equi-spaced within which $\Omega_\nu\in[\OmegaI, \OmegaII]$ is proposed here. The limits $\OmegaI$ and $\OmegaII$ are set similar to the maximum and minimum wave-frequencies where the attenuation model is to be valid. %

As power-law attenuation is commonly encountered for large wave-frequency intervals in complex media, this work puts emphasis on the determination of discrete relaxation parameters to yield power-law attenuation within a given frequency interval. %
The theoretical considerations of Section~\ref{sec:Theory} are supplemented by two numerical examples in Section~\ref{sec:numerical_example}. %
For the first example, where the attenuation is displayed in Fig.~\ref{fig:attenuation_model_parameter}, we note that for 3 relaxation the chosen weighted mechanisms give rise to an attenuation differs by less than $14\%$ relative to the pure power-law $\alpha_k(\omega) \propto \omega^{1.1}$ within the whole frequency region of interest $f\in[100\;\text{kHz}, 30\;\text{MHz}]$. %
For two mechanisms, the highest relative difference is $63\%$. 
By contrast, the \cite{Yang2005} weighted sum of three relaxations corresponds to a maximum relative difference in $\alpha_k(\omega)$ of $68\%$. When normalizing the Yang and Cleveland attenuation, the maximum difference is reduced to $29\%$, which is however still more than double the $14\%$ limit attained when setting the parameters following the prescription suggested in the current paper for the same number of relaxation mechanisms. %
Judging from the maximum relative absolute difference between the resulting $\alpha_k(\omega)$ and the pure power-law with the exponent $1.1$, the proposed method hence is more advantageous. %

\begin{figure}[t!]
	\begin{center}
		\insfig{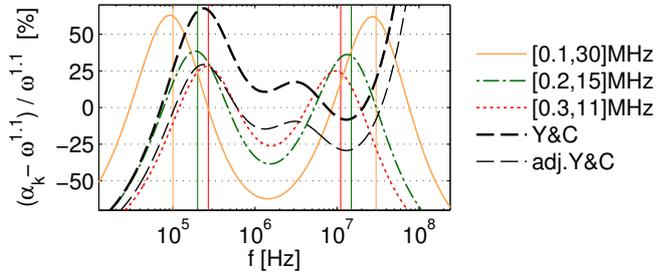}
	\end{center}
\caption{\label{fig:attenuation_YC_compareto_adjusted}(Color online)  %
Relative difference between the attained attenuation and the wanted power-law $\propto \omega^{1.1}$ for $N=2$ mechanisms, as attained when the relaxation frequencies are $\Omega_{1,2} = \{0.1,30\}$\:MHz (solid line), $\{0.2,15\}$\:MHz (dash-dotted line), and $\{0.3,11\}$\:MHz (dotted line). All these attenuation functions are normalized so that absolute maximum and absolute minimum of the relative difference are equal. For reference, the \cite{Yang2005} attenuation is also included, both in the original form (thick dashed line) and in the adjusted normalized form (thin dashed line). %
The vertical lines indicate frequencies of the explored relaxation mechanisms. %
} 
\end{figure}

Inspection of the resulting attenuation curves of Fig.~\ref{fig:attenuation_model_parameter} hints the possibility of attaining less relative difference between $\alpha_k(\omega)$ and the power-law by straightforward adjustment of the width of the populated relaxation frequency region. %
This possibility is now briefly explored. %
For low $N$, it seems appropriate to make the populated region smaller, while for larger $N$ the relative difference between attained $\alpha_k(\omega)$ and the wanted power-law increases close to $\OmegaI$ and $\OmegaII$ hence suggesting to make the populated frequency region wider. %
For $N=2$, shrinking the populated relaxation frequency range from $\Omega\in[0.1, 30]$\:MHz via $[0.2, 15]$\:MHz down to $[0.3,11]$\:MHz decreases the maximum relative difference from $63\%$ via $39\%$ down to $28\%$ (see Fig.~\ref{fig:attenuation_YC_compareto_adjusted}), which is actually about the same as was attained for the  normalized Yang and Cleveland fit for three relaxation processes. %

The second numerical example for which the resulting attenuation functions are displayed in Fig.~\ref{fig:attenuation_model_wid}, illustrates that that for as few as five discrete relaxation processes, %
an adequate power-law differing by less than $11\%$ from the wanted power-law may be constructed within the very wide frequency band $f\in[100\:\text{kHz}, 1\:\text{GHz}]$. %

The fractional Zener compressibility of Eq.~\eqref{eq:fZener_compressibility} is for $\alpha=\beta$ similar to the Cole--Cole expression for complex dielectric permittivity, which is empirically shown to be valid in a variety of complex media. %
A single discrete acoustical NSW relaxation mechanism in the dielectric permittivity context corresponds to a single Debye term. %
The adjustment and weighting of multiple discrete Debye terms to emulate Cole--Cole dielectrical behavior is treated \eg{} in \cite{Rekanos2010multiplerelax}, where a Pad\'e approximant approach is applied, in \cite{Tofighi2009} where an error-minimization method is employed choose the relaxation times at which the Cole--Cole relaxation distribution is sampled, in \cite{Kelley2007} where the relaxation frequencies are found using a nonlinear method and the weights are found using a linear least squares approach, as well as in \cite{Clegg2010} where a genetic algorithm is applied to find the multiple Debye parameters. %
The parameter determination approach brought forward in the present paper should also be tested out for selection of a discrete set of Debye terms to emulate the Cole--Cole dielectric permittivity. %

The model applied in \cite{Berkhoff1996} was contextualized in the discussion of Näsholm and Holm (2011). It was found that it corresponds to band-limited continuous relaxations with compressibility contributions given by $\kappa_\nu(\Omega) \propto \Omega^{-1}$. 

We note that based on experimental evidence, \cite{Jongen1986} suggests the attenuation for example in beef liver to be $\propto \omega^2$ below a certain cut-off frequency $\omega_c$ and $\propto \omega$ at higher frequencies. %
Such behavior is attained by the band-limited continuous relaxation distribution framework of the present work when the lowest frequency $\OmegaI$ of the populated relaxation frequency band is equal to $\omega_c$. %

The $N$ relaxation parameter determination method of Section~\ref{sec:discrete_NSW_bandlim_atten} corresponds to mapping a time-fractional wave equation into an integer-order one of highest order $N+2$ with the intent of the mapping being adequately valid within a given wave-frequency band. %
%
Because fractional-order differential wave equations require extra care in simulations, the method suggested here hence has a potential to facilitate numerical wave-propagation calculations. Comparison to previously suggested such numerical schemes as \eg{} \cite{Wismer1995, Caputo2011}, is an appropriate future work connected to this paper. %

From a practical point of view the method of this paper to determine discrete relaxation parameters is appealing because the anomalous physics of fractional-order differential equations is converted to a differential equation with an finite set of higher-order integer derivatives. %
This conversion is tailored for the frequency bandwidth of interest. %
For such conversions to be valid for all frequencies, an infinite number of terms are needed in the integer-order differential equation. Thence the narrowing of the frequency region is traded off into the convenience of having a finite highest derivative order. %

Finally {this work }calls upon %
further experimental verifications and more profound theoretical connection between the relaxation models and fundamental physical properties of complex materials. %
Hopefully the links between fractional calculus, observed attenuation behaviors, the stunningly common power-law patterns of nature, and the micromechanical structure of matter are to be further clarified. %



\begin{acknowledgments}
The author would like to thank Prof.~Sverre Holm for interesting discussions and valuable advice. %
This research was partly supported by the ``High Resolution Imaging and Beamforming'' project of the Norwegian Research Council.
\end{acknowledgments}

\appendix

%

\begin{thebibliography}{35}
\newcommand{\enquote}[1]{``#1''}
\expandafter\ifx\csname natexlab\endcsname\relax\def\natexlab#1{#1}\fi
\expandafter\ifx\csname url\endcsname\relax
  \def\url#1{\texttt{#1}}\fi
\expandafter\ifx\csname urlprefix\endcsname\relax\def\urlprefix{URL }\fi
\providecommand{\bibinfo}[2]{#2}
\providecommand{\noopsort}[1]{}
\providecommand{\switchargs}[2]{#2#1}

\bibitem[{Adolfsson \emph{et~al.}(2005)Adolfsson, Enelund, and
  Olsson}]{Adolfsson2005}
\bibinfo{author}{Adolfsson, K.}, \bibinfo{author}{Enelund, M.}, and
  \bibinfo{author}{Olsson, P.} (\textbf{\bibinfo{year}{2005}}).
  \enquote{\bibinfo{title}{On the fractional order model of viscoelasticity}},
  \bibinfo{journal}{Mech. Time-Dep. Mater.} \textbf{\bibinfo{volume}{9}},
  \bibinfo{pages}{15--34}.

\bibitem[{Ainslie and McColm(1998)}]{Ainslie1998}
\bibinfo{author}{Ainslie, M.} and \bibinfo{author}{McColm, J.~G.}
  (\textbf{\bibinfo{year}{1998}}). \enquote{\bibinfo{title}{{A simplified
  formula for viscous and chemical absorption in sea water}}},
  \bibinfo{journal}{J.\ Acoust.\ Soc.\ Am.} \textbf{\bibinfo{volume}{103}},
  \bibinfo{pages}{1671--1672}.

\bibitem[{Bagley and Torvik(1983)}]{Bagley83A}
\bibinfo{author}{Bagley, R.~L.} and \bibinfo{author}{Torvik, P.~J.}
  (\textbf{\bibinfo{year}{1983}}). \enquote{\bibinfo{title}{Fractional calculus
  --- {A} different approach to the analysis of viscoelastically damped
  structures}}, \bibinfo{journal}{AIAA J.} \textbf{\bibinfo{volume}{21}},
  \bibinfo{pages}{741--748}.

\bibitem[{Bass \emph{et~al.}(1995)Bass, Sutherland, Zuckerwar, Blackstock, and
  Hester}]{Bass1995}
\bibinfo{author}{Bass, H.}, \bibinfo{author}{Sutherland, L.},
  \bibinfo{author}{Zuckerwar, A.}, \bibinfo{author}{Blackstock, D.}, and
  \bibinfo{author}{Hester, D.} (\textbf{\bibinfo{year}{1995}}).
  \enquote{\bibinfo{title}{{Atmospheric absorption of sound: Further
  developments}}}, \bibinfo{journal}{J.\ Acoust.\ Soc.\ Am.}
  \textbf{\bibinfo{volume}{97}}, \bibinfo{pages}{680--683}.

\bibitem[{Bateman and Erd{\'e}lyi(1954)}]{bateman1954V1C6}
\bibinfo{author}{Bateman, H.} and \bibinfo{author}{Erd{\'e}lyi, A.}
  (\textbf{\bibinfo{year}{1954}}). \emph{\bibinfo{title}{Tables of integral
  transforms}}, volume~\bibinfo{volume}{1}, chapter~\bibinfo{chapter}{6}
  (\bibinfo{publisher}{McGraw--Hill}, \bibinfo{address}{New York}).

\bibitem[{Berkhoff \emph{et~al.}(1996)Berkhoff, Thijssen, and
  Homan}]{Berkhoff1996}
\bibinfo{author}{Berkhoff, A.~P.}, \bibinfo{author}{Thijssen, J.~M.}, and
  \bibinfo{author}{Homan, R. J.~F.} (\textbf{\bibinfo{year}{1996}}).
  \enquote{\bibinfo{title}{Simulation of ultrasonic imaging with linear arrays
  in causal absorptive media}}, \bibinfo{journal}{Ultrasound Med. Biol.}
  \textbf{\bibinfo{volume}{22}}, \bibinfo{pages}{245--259}.

\bibitem[{Caputo \emph{et~al.}(2011)Caputo, Carcione, and
  Cavallini}]{Caputo2011}
\bibinfo{author}{Caputo, M.}, \bibinfo{author}{Carcione, J.~M.}, and
  \bibinfo{author}{Cavallini, F.} (\textbf{\bibinfo{year}{2011}}).
  \enquote{\bibinfo{title}{Wave simulation in biologic media based on the
  {K}elvin--{V}oigt fractional-derivative stress--strain relation}},
  \bibinfo{journal}{Ultrasound Med. Biol.} \textbf{\bibinfo{volume}{37}},
  \bibinfo{pages}{996--1004}.

\bibitem[{Clegg and Robinson(2010)}]{Clegg2010}
\bibinfo{author}{Clegg, J.} and \bibinfo{author}{Robinson, M.~P.}
  (\textbf{\bibinfo{year}{2010}}). \enquote{\bibinfo{title}{A genetic algorithm
  used to fit {D}ebye functions to the dielectric properties of tissues}}, in
  \emph{\bibinfo{booktitle}{IEEE Congress on Evolutionary Computation}},
  \bibinfo{pages}{1--8}.

\bibitem[{Gl\"ockle and Nonnenmacher(1991)}]{Glockle1991}
\bibinfo{author}{Gl\"ockle, W.~G.} and \bibinfo{author}{Nonnenmacher, T.~F.}
  (\textbf{\bibinfo{year}{1991}}). \enquote{\bibinfo{title}{Fractional integral
  operators and {F}ox functions in the theory of viscoelasticity}},
  \bibinfo{journal}{Macromolecules} \textbf{\bibinfo{volume}{24}},
  \bibinfo{pages}{6426--6434}.

\bibitem[{Holm and N\"asholm(2011)}]{Holm2011}
\bibinfo{author}{Holm, S.} and \bibinfo{author}{N\"asholm, S.~P.}
  (\textbf{\bibinfo{year}{2011}}). \enquote{\bibinfo{title}{A causal and
  fractional all-frequency wave equation for lossy media}},
  \bibinfo{journal}{J.\ Acoust.\ Soc.\ Am.} \textbf{\bibinfo{volume}{130}},
  \bibinfo{pages}{2195--2202}.

\bibitem[{Holm and Sinkus(2010)}]{Holm2010}
\bibinfo{author}{Holm, S.} and \bibinfo{author}{Sinkus, R.}
  (\textbf{\bibinfo{year}{2010}}). \enquote{\bibinfo{title}{{A unifying
  fractional wave equation for compressional and shear waves}}},
  \bibinfo{journal}{J.\ Acoust.\ Soc.\ Am.} \textbf{\bibinfo{volume}{127}},
  \bibinfo{pages}{542--548}.

\bibitem[{Jongen \emph{et~al.}(1986)Jongen, Thijssen, van~den Aarssen, and
  Verhoef}]{Jongen1986}
\bibinfo{author}{Jongen, H. A.~H.}, \bibinfo{author}{Thijssen, J.~M.},
  \bibinfo{author}{van~den Aarssen, M.}, and \bibinfo{author}{Verhoef, W.~A.}
  (\textbf{\bibinfo{year}{1986}}). \enquote{\bibinfo{title}{A general model for
  the absorption of ultrasound by biological tissues and experimental
  verification}}, \bibinfo{journal}{J.\ Acoust.\ Soc.\ Am.}
  \textbf{\bibinfo{volume}{79}}, \bibinfo{pages}{535--540}.

\bibitem[{Kelley \emph{et~al.}(2007)Kelley, Destan, and Luebbers}]{Kelley2007}
\bibinfo{author}{Kelley, D.~F.}, \bibinfo{author}{Destan, T.~J.}, and
  \bibinfo{author}{Luebbers, R.~J.} (\textbf{\bibinfo{year}{2007}}).
  \enquote{\bibinfo{title}{Debye function expansions of complex permittivity
  using a hybrid particle swarm-least squares optimization approach}},
  \bibinfo{journal}{IEEE Trans. Antennas Propag.}
  \textbf{\bibinfo{volume}{55}}, \bibinfo{pages}{1999--2005}.

\bibitem[{Kelly and McGough(2009)}]{Kelly2009}
\bibinfo{author}{Kelly, J.~F.} and \bibinfo{author}{McGough, R.~J.}
  (\textbf{\bibinfo{year}{2009}}). \enquote{\bibinfo{title}{Fractal ladder
  models and power law wave equations}}, \bibinfo{journal}{J.\ Acoust.\ Soc.\
  Am.} \textbf{\bibinfo{volume}{126}}, \bibinfo{pages}{2072--2081}.

\bibitem[{Liebler \emph{et~al.}(2004)Liebler, Ginter, Dreyer, and
  Riedlinger}]{Liebler2004}
\bibinfo{author}{Liebler, M.}, \bibinfo{author}{Ginter, S.},
  \bibinfo{author}{Dreyer, T.}, and \bibinfo{author}{Riedlinger, R.~E.}
  (\textbf{\bibinfo{year}{2004}}). \enquote{\bibinfo{title}{Full wave modeling
  of therapeutic ultrasound: {E}fficient time-domain implementation of the
  frequency power-law attenuation}}, \bibinfo{journal}{J.\ Acoust.\ Soc.\ Am.}
  \textbf{\bibinfo{volume}{116}}, \bibinfo{pages}{2742--2750}.

\bibitem[{Nachman \emph{et~al.}(1990)Nachman, Smith~III, and
  Waag}]{Nachman1990}
\bibinfo{author}{Nachman, A.~I.}, \bibinfo{author}{Smith~III, J.~F.}, and
  \bibinfo{author}{Waag, R.~C.} (\textbf{\bibinfo{year}{1990}}).
  \enquote{\bibinfo{title}{An equation for acoustic propagation in
  inhomogeneous media with relaxation losses}}, \bibinfo{journal}{J.\ Acoust.\
  Soc.\ Am.} \textbf{\bibinfo{volume}{88}}, \bibinfo{pages}{1584--1595}.

\bibitem[{N\"asholm and Holm(2011)}]{Nasholm2011}
\bibinfo{author}{N\"asholm, S.~P.} and \bibinfo{author}{Holm, S.}
  (\textbf{\bibinfo{year}{2011}}). \enquote{\bibinfo{title}{Linking multiple
  relaxation, power-law attenuation, and fractional wave equations}},
  \bibinfo{journal}{J.\ Acoust.\ Soc.\ Am.} \textbf{\bibinfo{volume}{130}},
  \bibinfo{pages}{3038--3045}.

\bibitem[{Newman(2005)}]{newman2005}
\bibinfo{author}{Newman, M. E.~J.} (\textbf{\bibinfo{year}{2005}}).
  \enquote{\bibinfo{title}{Power laws, {P}areto distributions and {Z}ipf's
  law}}, \bibinfo{journal}{Contemp. Phys} \textbf{\bibinfo{volume}{46}},
  \bibinfo{pages}{323--351}.

\bibitem[{Papoulia \emph{et~al.}(2010)Papoulia, Panoskaltsis, Kurup, and
  Korovajchuk}]{Papoulia2010}
\bibinfo{author}{Papoulia, K.}, \bibinfo{author}{Panoskaltsis, V.},
  \bibinfo{author}{Kurup, N.}, and \bibinfo{author}{Korovajchuk, I.}
  (\textbf{\bibinfo{year}{2010}}). \enquote{\bibinfo{title}{Rheological
  representation of fractional order viscoelastic material models}},
  \bibinfo{journal}{Rheol. Acta} \textbf{\bibinfo{volume}{49}},
  \bibinfo{pages}{381--400}.

\bibitem[{Pauly and Schwan(1971)}]{Pauly1971}
\bibinfo{author}{Pauly, H.} and \bibinfo{author}{Schwan, H.~P.}
  (\textbf{\bibinfo{year}{1971}}). \enquote{\bibinfo{title}{Mechanism of
  absorption of ultrasound in liver tissue}}, \bibinfo{journal}{J.\ Acoust.\
  Soc.\ Am.} \textbf{\bibinfo{volume}{50}}, \bibinfo{pages}{692--699}.

\bibitem[{Prieur and Holm(2011)}]{Prieur2011}
\bibinfo{author}{Prieur, F.} and \bibinfo{author}{Holm, S.}
  (\textbf{\bibinfo{year}{2011}}). \enquote{\bibinfo{title}{Nonlinear acoustic
  wave equations with fractional loss operators}}, \bibinfo{journal}{J.\
  Acoust.\ Soc.\ Am.} \textbf{\bibinfo{volume}{130}},
  \bibinfo{pages}{1125--1132}.

\bibitem[{Prieur \emph{et~al.}(2012)Prieur, Vilenskiy, and Holm}]{Prieur2012}
\bibinfo{author}{Prieur, F.}, \bibinfo{author}{Vilenskiy, G.}, and
  \bibinfo{author}{Holm, S.} (\textbf{\bibinfo{year}{2012}}).
  \enquote{\bibinfo{title}{A more fundamental approach to the derivation of
  nonlinear acoustic wave equations with fractional loss operators}},
  \bibinfo{journal}{J.\ Acoust.\ Soc.\ Am.} \textbf{\bibinfo{volume}{132}},
  \bibinfo{pages}{2169--2172}.

\bibitem[{Rekanos and Papadopoulos(2010)}]{Rekanos2010multiplerelax}
\bibinfo{author}{Rekanos, I.~T.} and \bibinfo{author}{Papadopoulos, T.~G.}
  (\textbf{\bibinfo{year}{2010}}). \enquote{\bibinfo{title}{{FDTD} modeling of
  wave propagation in {C}ole--{C}ole media with multiple relaxation times}},
  \bibinfo{journal}{IEEE Antenn. Wireless Propag. Lett.}
  \textbf{\bibinfo{volume}{9}}, \bibinfo{pages}{67--69}.

\bibitem[{Roitner \emph{et~al.}(2012)Roitner, Bauer-Marschallinger, Berer, and
  Burgholzer}]{roitner2012experimental}
\bibinfo{author}{Roitner, H.}, \bibinfo{author}{Bauer-Marschallinger, J.},
  \bibinfo{author}{Berer, T.}, and \bibinfo{author}{Burgholzer, P.}
  (\textbf{\bibinfo{year}{2012}}). \enquote{\bibinfo{title}{Experimental
  evaluation of time domain models for ultrasound attenuation losses in
  photoacoustic imaging}}, \bibinfo{journal}{J.\ Acoust.\ Soc.\ Am.}
  \textbf{\bibinfo{volume}{131}}, \bibinfo{pages}{3763--3774}.

\bibitem[{Rossikhin and Shitikova(2001)}]{rossikhin2001analysis}
\bibinfo{author}{Rossikhin, Y.~A.} and \bibinfo{author}{Shitikova, M.~V.}
  (\textbf{\bibinfo{year}{2001}}). \enquote{\bibinfo{title}{Analysis of
  rheological equations involving more than one fractional parameters by the
  use of the simplest mechanical systems based on these equations}},
  \bibinfo{journal}{Mech. Time-Depend. Mat.} \textbf{\bibinfo{volume}{5}},
  \bibinfo{pages}{131--175}.

\bibitem[{Samko \emph{et~al.}(1993)Samko, Kilbas, and
  Marichev}]{samko1993chapter2}
\bibinfo{author}{Samko, S.~G.}, \bibinfo{author}{Kilbas, A.~A.}, and
  \bibinfo{author}{Marichev, O.~I.} (\textbf{\bibinfo{year}{1993}}).
  \emph{\bibinfo{title}{Fractional Integrals and Derivatives: Theory and
  Applications}}, chapter~\bibinfo{chapter}{2} (\bibinfo{publisher}{Gordon and
  Breach}, \bibinfo{address}{New York}).

\bibitem[{Szabo and Wu(2000)}]{Szabo00}
\bibinfo{author}{Szabo, T.~L.} and \bibinfo{author}{Wu, J.}
  (\textbf{\bibinfo{year}{2000}}). \enquote{\bibinfo{title}{A model for
  longitudinal and shear wave propagation in viscoelastic media}},
  \bibinfo{journal}{J.\ Acoust.\ Soc.\ Am.} \textbf{\bibinfo{volume}{107}},
  \bibinfo{pages}{2437--2446}.

\bibitem[{Tabei \emph{et~al.}(2003)Tabei, Mast, and Waag}]{Tabei2003}
\bibinfo{author}{Tabei, M.}, \bibinfo{author}{Mast, T.~D.}, and
  \bibinfo{author}{Waag, R.~C.} (\textbf{\bibinfo{year}{2003}}).
  \enquote{\bibinfo{title}{Simulation of ultrasonic focus aberration and
  correction through human tissue}}, \bibinfo{journal}{J.\ Acoust.\ Soc.\ Am.}
  \textbf{\bibinfo{volume}{113}}, \bibinfo{pages}{1166--1176}.

\bibitem[{Tofighi(2009)}]{Tofighi2009}
\bibinfo{author}{Tofighi, M.-R.} (\textbf{\bibinfo{year}{2009}}).
  \enquote{\bibinfo{title}{{FDTD} modeling of biological tissues {C}ole--{C}ole
  dispersion for 0.5--30 {GHz} using relaxation time distribution samples ---
  novel and improved implementations}}, \bibinfo{journal}{IEEE Trans. Microw.
  Theory Tech.} \textbf{\bibinfo{volume}{57}}, \bibinfo{pages}{2588--2596}.

\bibitem[{Treeby \emph{et~al.}(2012)Treeby, Jaros, Rendell, and
  Cox}]{treeby2012modeling}
\bibinfo{author}{Treeby, B.~E.}, \bibinfo{author}{Jaros, J.},
  \bibinfo{author}{Rendell, A.~P.}, and \bibinfo{author}{Cox, B.~T.}
  (\textbf{\bibinfo{year}{2012}}). \enquote{\bibinfo{title}{Modeling nonlinear
  ultrasound propagation in heterogeneous media with power law absorption using
  a $k$-space pseudospectral method}}, \bibinfo{journal}{J.\ Acoust.\ Soc.\
  Am.} \textbf{\bibinfo{volume}{131}}, \bibinfo{pages}{4324--4336}.

\bibitem[{Vilensky \emph{et~al.}(2012)Vilensky, ter Haar, and
  Saffari}]{Vilensky2012}
\bibinfo{author}{Vilensky, G.}, \bibinfo{author}{ter Haar, G.}, and
  \bibinfo{author}{Saffari, N.} (\textbf{\bibinfo{year}{2012}}).
  \enquote{\bibinfo{title}{A model of acoustic absorption in fluids based on a
  continuous distribution of relaxation times}}, \bibinfo{journal}{Wave Motion}
  \textbf{\bibinfo{volume}{49}}, \bibinfo{pages}{93--108}.

\bibitem[{Widder(1966)}]{widder1966transform}
\bibinfo{author}{Widder, D.~V.} (\textbf{\bibinfo{year}{1966}}).
  \enquote{\bibinfo{title}{A transform related to the poisson integral for a
  half-plane}}, \bibinfo{journal}{Duke Math. J.} \textbf{\bibinfo{volume}{33}},
  \bibinfo{pages}{355--362}.

\bibitem[{Wismer(2006)}]{Wismer06}
\bibinfo{author}{Wismer, M.~G.} (\textbf{\bibinfo{year}{2006}}).
  \enquote{\bibinfo{title}{Finite element analysis of broadband acoustic pulses
  through inhomogenous media with power law attenuation}},
  \bibinfo{journal}{J.\ Acoust.\ Soc.\ Am.} \textbf{\bibinfo{volume}{120}},
  \bibinfo{pages}{3493--3502}.

\bibitem[{Wismer and Ludwig(1995)}]{Wismer1995}
\bibinfo{author}{Wismer, M.~G.} and \bibinfo{author}{Ludwig, R.}
  (\textbf{\bibinfo{year}{1995}}). \enquote{\bibinfo{title}{An explicit
  numerical time domain formulation to simulate pulsed pressure waves in
  viscous fluids exhibiting arbitrary frequency power law attenuation}},
  \bibinfo{journal}{IEEE Trans.\ Ultrason.\ Ferroelectr.,\ Freq.\ Control}
  \textbf{\bibinfo{volume}{42}}, \bibinfo{pages}{1040--1049}.

\bibitem[{Yang and Cleveland(2005)}]{Yang2005}
\bibinfo{author}{Yang, X.} and \bibinfo{author}{Cleveland, R.~O.}
  (\textbf{\bibinfo{year}{2005}}). \enquote{\bibinfo{title}{Time domain
  simulation of nonlinear acoustic beams generated by rectangular pistons with
  application to harmonic imaging}}, \bibinfo{journal}{J.\ Acoust.\ Soc.\ Am.}
  \textbf{\bibinfo{volume}{117}}, \bibinfo{pages}{113--123}.

\end{thebibliography}

\end{document}